\documentclass[11pt,b5paper,twoside]{article}      

\newcommand{\mechyear}{2017}
\newcommand{\mechvol}{31}

\usepackage[cp1251]{inputenc}
\usepackage[T2A]{fontenc}
\usepackage[russian]{babel}

\usepackage[dvips]{graphicx}
\usepackage[dvips]{color}

\usepackage{amsmath}
\usepackage{amssymb}
\usepackage{amsxtra}
\usepackage{latexsym}
\usepackage{ifthen}
\usepackage[centerlast,footnotesize,sl]{caption2}
\usepackage{bm}

\usepackage{trudyn}

\usepackage{amsmath}
\usepackage{graphicx}

\newcommand{\Zed}{\mathbb Z}

\begin{document}


\udk{531.38}

\title{О КО-КОНТР-ФРАГМЕНТАХ ПОВЕДЕНИЯ АВТОМАТОВ}
      {О ко-контр-фрагментах поведения автоматов}


\titleeng{On co-counter-fragments of behaviour of automata}


\author{А.Н. Курганский, А.Ю. Максимова}
\authoreng{O. Kurganskyy, A.J. Maksimova}

\date{27.11.2017}




\address{ГУ <<ИПММ>>, Донецк}


\email{kurgansk@iamm.su}

\maketitle

\begin{abstract}
Работа относится к области теории восстановления и представления абстрактных автоматов фрагментами поведения и посвящена исследованию структуры классов конечных связных инициальных автоматов без выхода, задаваемых системами определяющих соотношений, рассматриваемых в роли фрагментов, кофрагментов, контрфрагментов и коконтрфрагментов автоматов.
\vspace{1mm}\\
\textbf{\emph {Ключевые слова: }}\itshape{фрагменты, кофрагменты, контрфрагменты, коконтрфрагменты, системы определяющих соотношений, представление автоматов}
\end{abstract}

\abstracteng{This paper contains results related to synthesis and presentation of abstract automata by fragments of behaviour and investigates the structure of the classes of finite connected initial output-less automata specified by systems of defining relations considered as fragments, co-fragments, counter-fragments and co-counter-fragments of automata.}


\kweng{fragments, co-fragments, counter-fragments, co-counter-fragments, systems of defining relations, presentation of automata}

\section{Введение}

Одно из современных направлений развития теории автоматов отталкивается от понятия коалгебры, которое является двойственным понятию алгебры и рассматривается как определение динамических систем на языке теории категорий~\cite{Sangiorgi2012-I,Sangiorgi2012-A,Rutten2000}. Данному направлению развития принадлежат такие понятия как коиндукция и бисимуляция.
Настоящая работа лежит в области теории представления и восстановления автоматов по фрагментам поведения, разработанной в~\cite{GruKozPon,Gru2003,GruKoz2004}, и является развитием на основе идей коалгебраического подхода работы~\cite{GruSen2004}, посвященной системам определяющих соотношений конечных связных инициальных автоматов без выхода. 
Современное состояние теория представления автоматов фрагментами поведения можно найти в~\cite{KudGruKoz2009-1,KudGruKoz2009-2,KudGruKoz2009-3}.

Инициальный автомат порождает отношение на словах, называемое правой конгруэнцией автомата, по правилу: слова находятся в отношении правой конгруэнции автомата тогда и только тогда, когда они ведут в одно и то же состояние из начального. Произвольное множество пар слов называется определяющими соотношениями автомата, если правоконгруэнтное замыкание этого множества совпадает с правой конгруэнцией автомата. 
С точки зрения теории экспериментов с автоматами элементы правой конгруэнции автомата являются фрагментами его поведения и структуры. В теории автоматов фундаментальна проблема минимизации автомата по числу состояний. Однако правоконгруэнтное замыкание отношения на словах дает автоматы с максимальным числом состояний из всех тех, для которых эти отношения выполняются. Это говорит о том, что определяющие соотношения и их конгруэнтное замыкание скорее связаны со структурой графа автомата, чем с его поведением, и, следовательно, не вполне соответствуют поведенческой проблематике теории автоматов. Цель данной работы рассмотреть  элементы правой конгруэнции не только в роли фрагментов автомата, но и в роли определяемых в работе  кофрагментов, контрфрагментов и коконтрфрагментов (короче: (ко)\-(контр)\-фраг\-ментов), что позволит связать определяющие отношения как со структурным, так и поведенческим контекстом анализа и синтеза автоматов.
Можно сказать, что фрагменты, кофрагменты, контрфрагменты и коконтрфрагменты автоматов преставляют собой, соответственно, необходимое, разрешенное, запрещенное как противоречие и запрещенное как контрпример поведение.
Точка зрения на фрагменты в роли (ко)\-(контр)\-фраг\-ментов лежит в русле идей работы~\cite{KudGruKoz2009-2} и является развитием некоторых её моментов.

\section{Постановка задачи}

Пусть дан произвольный класс автоматов, который назовём  \textit{допустимым}.  
Дано задание в виде (к примеру: теоретико-множественной) формулы над  (ко)\-(контр)\-фраг\-ментами на восстановление неизвестного автомата из допустимого класса.  Условие задания в виде формулы $F$ над (ко)\-(контр)\-фраг\-ментами назовем \textit{представлением} автомата. Автоматы, являющиеся решениями этого задания, назовем  \textit{совместимыми} с представлением $F$. Множество всех совместимых с представлением автоматов назовем \textit{объемом} представления.
\begin{definition}
Представление называет непротиворечивым, если его объем не пуст, и называется полным, если его объем состоит из одного автомата (с принятой к рассмотрению точностью, например, с точностью до изоморфизма или бисимуляции). В противном случае, представление называется противоречивым.
\end{definition}

Цель данной работы ввести понятия  (ко)\-(контр)\-фраг\-мента автомата и исследовать в частных случаях структуру объема представления. 

\section{Общие замечания}
\subsection{Логика восстановления автоматов}
Представление автомата (ко)(контр)фрагментами формируется на основе спецификации (априорной информации) и экспериментов с автоматом. Оно содержит в неявном виде частичную или полную информацию об автомате. Синтез автомата по этой информации подразумевает некоторую логику получения новой явной информации о поведении и структуре автомата. Логика восстановления автомата позволяет получать (перечислять) новые факты об автомате. В пределе с помощью некоторого алгоритма она дает полную информацию в готовом замкнутом виде, например, таблице автомата. Таким образом, проводя аналогии с перечислимыми и рекурсивными (разрешимыми) множествами, можно сказать, что процесс восстановления автомата есть переход от неявной формы представления автомата, позволяющей без дополнительных вычислительных усилий лишь перечислять факты о его поведении, к явной, рекурсивной, форме задания автомата.

В работе мы проводим следующие аналогии между синтаксическими теориями в логике и конечными автоматами, рассматриваемых в контексте систем определяющих соотношений. Синтаксическая теория представляет собой множество правильно построенных утверждений (формул), среди которых выделены аксиомы, и правил вывода. Правила вывода базируясь на аксиомах разбивают формулы на два класса: выводимые формулы (доказуемые теоремы) и невыводимые формулы. Собственно, теоремы и образуют саму теорию. Аксиомы и всякое множество теорем являются фрагментами теории, то есть частью, абстракцией теории. Аксиомы в неявном виде логически содержат всю теорию. Теоремы в неявном содержат аксиомы. При изменении множества аксиом меняется теория. Мы придерживаемся аналогий между: 1) правильно построенными утверждениями и (ко)\-(контр)\-фраг\-ментами, 2) теорией и автоматом, 3) построением теории и синтезом автомата, 4) аксиомами теории и фрагментами автомата. Эти аналогии в свою очередь влекут аналогии между: 1) теоремами теории и фрагментами автомата, 2) моделями теории и кофрагментами, 2) противоречиями и контрфрагментами, 3) контрпримерами в доказательствах и коконтрфрагментами. Аналогии будут разъяснены ниже.
Теперь в общих определениях объясним смысл понятий (ко)\-(контр)\-фраг\-ментов. 

\subsection{Фрагменты}

Фрагмент автомата -- это информация о необходимом поведении и структуре автомата, это часть автомата, его абстракция. 
Фрагмент является абс\-трак\-т\-но-все\-об\-щим определением совместимого с ним класса автоматов.
 
 Восстановление автомата означает синтетическое пополнение содержания фрагмента, расширение его до некоторого допустимого и совместимого с фрагментом автомата. Этот процесс есть выполнение некоторого алгоритма. Результат действия алгоритма будем называть \textit{замыканием} фрагмента. Вообще говоря, замыкание фрагмента неоднозначно. Ниже будут выделены частные случаи: индуктивное и коиндуктивное замыкания фрагмента.
\begin{definition}
Фрагмент назовем замкнутым, если он является допустимым автоматом.
\end{definition}

\subsection{Решётка фрагментов и допустимых автоматов}
Вводимые ниже понятия (ко)(контр)фрагментов автомата отталкиваются от понятия фрагмента вообще.  Мы считаем, что дано множество объектов, называемых фрагментами, которое включает в себя и множество допустимых автоматов. Между фрагментами и допустимыми автоматами установлено отношение <<быть фрагментом>>. 
Если фрагмент $F$ является фрагментом автомата $A$, то пишем $F\le A$ и говорим также, что $A$ содержит фрагмент $F$ или $A$ \textit{совместим} с фрагментом $F$.  

Считаем, что допустимые автоматы с отношением $\le$ образует полную решётку с операциями $\sup$ и $\inf$, и что, кроме того, из $F\le A$, $F\le B$ следует $F\le\inf\{A,B\}$ для любых допустимых автоматов $A$, $B$ и фрагмента $F$. Такое упрощение сужает разнообразие принимаемых во внимание классов допустимых автоматов, но для цели данной работы, заключающейся в определении понятий (ко)(контр)фрагментов и представления с их помощью автоматов, обобщающих понятие фрагмента автомата, введенного в~\cite{GruKozPon,Gru2003,GruKoz2004}, данное ограничение удовлетворительно. Наибольший в решётке допустимый автомат обозначим через $[1]$, наименьший через $[0]$.

\begin{definition}
Индуктивным замыканием фрагмента $A$ называется допустимый автомат $[A]$, являющийся наименьшим в решётке среди всех допустимых автоматов, содержащих $A$.
\end{definition}

Введем отношение быть фрагментом на самих фрагментах.

\begin{definition}
Фрагмент $A$ является фрагментом фрагмента $B$, если индуктивное замыкание $[A]$  является фрагментом индуктивного замыкания  $[B]$. 
\end{definition}
Два фрагмента $A$, $B$ назовем эквивалентными, если $A\le B$ и $B\le A$. Решётка классов эквивалентности фрагментов изоморфна решётке допустимых автоматов. Соответствующий изоморфизм ставит в соответствие автомату класс эквивалентности фрагментов, содержащий этот автомат.

Обозначим класс допустимых автоматов, совместимых с фрагментом $F$,  через $\operatorname{Fr}(F)$. Класс $\operatorname{Fr}(F)$ является полной решёткой с наименьшим (отношение $\le$) автоматом $[F]$ и наибольшим $[1]$.

Автомат $[1]$ совместим с любым фрагментом, а в рассматриваемом ниже примере автомат $[1]$ имеет наименьшее число состояний из всех допустимых автоматов. Таким образом, представление автоматов с помощью фрагментов имеет вырожденное, тривиальное и, с точки зрения минимизации, наилучшее решение. Это указывает на несоответствие фрагментов авто\-мат\-но-пове\-ден\-чес\-кой проблематике при синтезе автоматов. Однако в задачах навигации и распознавания карт, зачастую задаваемых графами с размеченными вершинами или дугами, то есть как и автоматы, фрагменты играют другую роль, поскольку на первый план выдвигается скорее структура объекта, а не его поведение. 

\subsection{Кофрагменты = модели}

Кофрагмент автомата является двойственным понятием к фрагменту и описывает разрешенное поведение. Если $A$ является фрагментом $B$, то $B$ является кофрагментом $A$.
Кофрагмент является ограничением сверху на поведение или структуру восстанавливаемого автомата. Он описывает класс совместимых с ним автоматов: всякий автомат класса является фрагментом кофрагмента класса, то есть общим между автоматами класса является то, что они все являются частью одного и того же. 

В представленном ниже примере мы свяжем аналогией понятия кофрагмента и модели теории, а также -- контрпримера и коконтрфрагмента, противоречия и контрфрагмента.

\begin{definition}
Фрагмент $B$ является кофрагментом фрагмента $A$, если индуктивное замыкание фрагмента  $A$ (то есть некоторый допустимый автомат) является фрагментом индуктивного замыкания фрагмента  $B$. 
\end{definition}

Обозначим класс допустимых автоматов, для которых $F$ является кофрагментом через $\operatorname{co-Fr}(F)$. Класс $\operatorname{co-Fr}(F)$ является полной решёткой с наименьшим (отношение $\le$) автоматом $[0]$ и наибольшим $F$.

В математической логике моделью синтаксической теории $T$ является некоторая содержательная математическая структура, которая в свою очередь может быть аксиоматизирована некоторой синтаксической теорией $T'$, в которой аксиомы исходной теории $T$ являются теоремами. Таким образом, можно сказать, что моделью одной синтаксической теории $T$ является другая синтаксическая теория $T'$ такая, что аксиомы первой ($T$) являются теоремами во второй ($T'$). Теоремы модели играют роль содержательной истины для $T$, остальные формулы играют роль ложных утверждений.  На языке (ко)\-(контр)\-фрагментов в теории представления автоматов модель переходит в роль кофрагмента.  Поведение, не являющееся фрагментом кофрагмента автомата, играет роль лжи. Понятие лжи позволяет вводить понятие контрпримера (коконтрфрагмента) в том смысле, в котором его используют для доказательства ложности утверждений. 

\subsection{Контрфрагменты = противоречия}
Контрфрагмент -- запрещенное поведение или запрещенный элемент структуры автомата. Подобно тому, как теория считается противоречивой, если в ней выводятся противоречия, то есть некоторые заведомо недопустимые факты как, к примеру, $A=\bar{A}$, то и представление автомата противоречиво, если синтез на его основе дает автомат содержащий контрфрагмент.

Пусть $K$ класс допустимых автоматов. Обозначим через $\operatorname{contra-Fr}(A)$ класс допустимых автоматов, для которых $A$ не является фрагментом: 
$$\operatorname{contra-Fr}(A)=K-\operatorname{Fr}(A)=\overline{\operatorname{Fr}(A)}.$$

\subsection{Коконтрфрагменты = контрпримеры}

Коконтрфрагмент -- двойственное понятие к контрфрагменту: если $B$ коконтрфрагмент автомата $A$, то $A$ не является фрагментом $B$.

Обозначим через $\operatorname{co-contra-Fr}(A)$ класс допустимых автоматов, для которых $A$ не является кофрагментом: 
$$\operatorname{co-contra-Fr}(A)=K-\operatorname{co-Fr}(A)=\overline{\operatorname{co-Fr}(A)}.$$

Упомянутую выше аналогию между коконтрфрагментами и контрпримерами рассмотрим на примере поиска контрпримера для пропозициональной формулы в исчислении секвенций (подробности к теме поиска контрпримера в исчислении секвенций в~\cite{ShenVer}). Напомним, что секвенцией называется выражение вида $\Gamma\vdash\Delta$, где $\Gamma$ и $\Delta$~-- некоторые конечные множества формул исчисления высказываний. Если $A$ пропозициональная формула, то, в силу соответствующей теоремы о полноте, секвенция $\vdash A$ выводима тогда и только тогда, когда $A$ тавтология. Аксиомами исчисления секвенций являются секвенции, в которых $\Gamma$ и $\Delta$ представляют собой наборы переменных, причем пересечение $\Gamma$ и $\Delta$ непусто. Можно сказать, что $\vdash A$ является контрфрагментом теории, если $A$ не тавтология. Для доказательства того, что секвенция $\vdash A$ является теоремой исчисления, можно пойти двумя путями. Первый путь заключается в том, чтобы из этой секвенции и аксиом вывести контрфрагмент (противоречие), например, $\vdash (x\wedge\bar{x})$. Второй путь основывается на теореме о полноте, связывающей теорию и модель, и идёт, в некотором смысле, в другом направлении -- от теоремы к аксиомам модели, в направлении анализа формулы с целью найти булевы значения переменных, при которых формула как булева функция принимает значение $0$. Другими словами, надо найти такое множество секвенций, содержащих только переменные, из которых выводится $\vdash A$, и при этом хотя бы одна из секвенций не является аксиомой, то есть имеет вид $\Gamma\vdash\Delta$, в котором $\Gamma$ и $\Delta$ представляют собой наборы переменных и пересечение $\Gamma$ и $\Delta$ пусто. Тогда при значении $1$ переменных из $\Gamma$ и значении $0$ переменных из $\Delta$ формула $A$ принимает значение $0$. Вот такие множества секвенций из переменных, содержащих не только аксиомы, и являются аналогами коконтрфрагментов в представлении автоматов.

\begin{definition}
Фрагмент $A'$ называется контрфрагментом (коконтрфрагментом) фрагмента $A$, если 
$[A]\in\operatorname{contra-Fr}(A')$ (соответственно, $[A]\in\operatorname{co-contra-Fr}(A')$).
\end{definition}

В определении через $[A]$ обозначается индуктивное замыкание фрагмента $A$.

\subsection{Составные фрагменты}

Из фрагментов можно составлять сложные фрагменты. Если $A$ и $B$ фрагменты, то составной фрагмент будем обозначачть через $A+B$. Это может быть, в зависимости от природы фрагментов, либо объединение фрагментов как множеств, либо прямая сумма частичных автоматов и т.д. По определению считаем, что  
\[
\operatorname{Fr}(A+B)=\operatorname{Fr}(\sup(A,B))=\operatorname{Fr}(A)\cap\operatorname{Fr}(B).
\]
Отсюда сразу следует:
\[
\operatorname{contra-Fr}(A+B)=\operatorname{contra-Fr}(A)\cap\operatorname{contra-Fr}(B).
\]
Аналогично, по определению:
\[
\operatorname{co-Fr}(A+B)=\operatorname{co-Fr}(\inf(A,B))=\operatorname{Fr}(A)\cap\operatorname{Fr}(B).
\]
Откуда следует:
\[
\operatorname{co-contra-Fr}(A+B)=\operatorname{contra-Fr}(A)\cap\operatorname{contra-Fr}(B).
\]

\section{Определения}
Обозначим класс допустимых автоматов $K$. Пусть $M=\left(S,X,\delta,s_0\right)$ -- допустимый автомат с множеством состояний $S$, входным алфавитом $X$, функцией перехода $\delta:S\times X\to S$ и начальным состоянием $s_0\in S$. Автомат порождает на $X^*$ правую конгруэнцию $\rho_M$ по правилу $p\rho_M q$, если $s_0p=s_0q$, где, если $s$~-- состояние, а $p$~-- слово, то $sp=\delta(s,p)$. 

Пусть $\rho \subseteq X^*\times X^*$. Считаем, что существует такое минимальное по включению отношение $[\rho]$, $\rho\subseteq [\rho]$, что $[\rho]=\rho_M$ для некоторого допустимого автомата $M$. Назовём $[\rho]$ $K$-допустимым замыканием отношения $\rho$. К примеру, если бы допустимый класс содержал все автоматы в алфавите $X$, то $[\rho]$ было бы правоконгруэнтным замыканием отношения $\rho$.  Мы отождествляем автомат $M$ и его отношение правой конгруэнции $\rho_M$. 
\begin{definition}
Отношение $\rho$ называется фрагментом отношения $\rho'$, а $\rho'$ кофрагментом $\rho$, если $[\rho]\subseteq[\rho']$.
\end{definition}
Обозначим класс допустимых автоматов, для которых отношение $\rho$ является фрагментом (кофрагментом) через $\operatorname{Fr}(\rho)$ (соответственно, $\operatorname{co-Fr}(\rho)$).

Обозначим через $\operatorname{contra-Fr}(\rho)$ класс допустимых автоматов, для которых $\rho$ не является фрагментом, $\operatorname{contra-Fr}(\rho)=K-\operatorname{Fr}(\rho)=\overline{\operatorname{Fr}(\rho)}$. Аналогично, 
\[
\operatorname{co-contra-Fr}(\rho)=K-\operatorname{co-Fr}(\rho)=\overline{\operatorname{co-Fr}(\rho)}.
\]
\begin{definition}
Отношение $\rho'$ называется контрфрагментом (коконтрфрагментом) отношения $\rho$, если 
$[\rho]\in\operatorname{contra-Fr}(\rho')$ (соответственно, $[\rho]\in\operatorname{co-contra-Fr}(\rho')$).
\end{definition}
Иллюстрация идеи, вкладываемой в определения (ко)\-(контр)\-фраг\-мен\-тов, показана на рисунке.
\begin{figure}[!ht]
\centering
\includegraphics[width=1.0\textwidth]{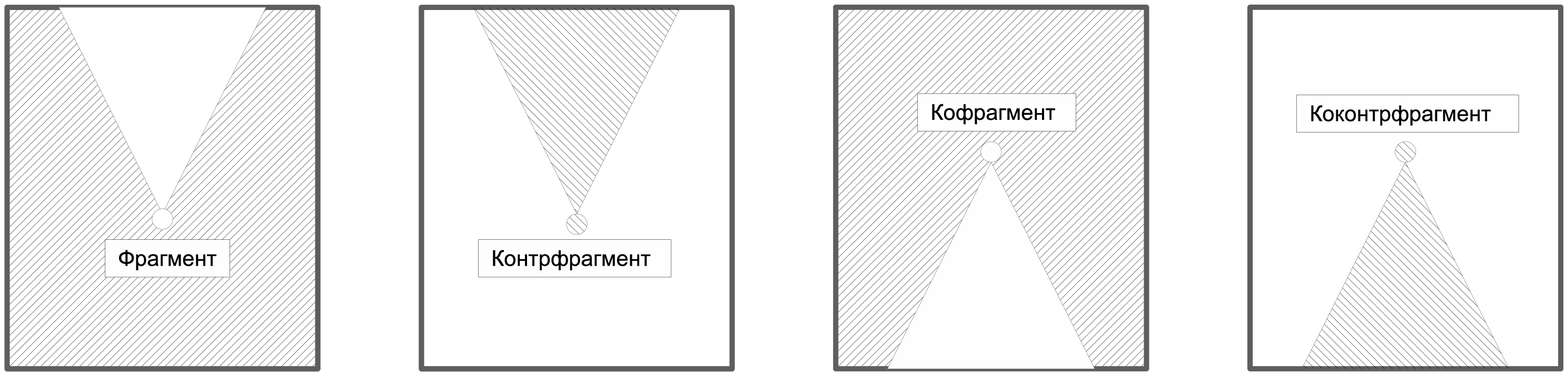}
\caption{Заштрихованная часть не содержит автоматов определяемого класса}
\end{figure}

Множества вида $\operatorname{Fr}(\rho)$, $\operatorname{co-Fr}(\rho)$, $\operatorname{contra-Fr}(\rho)$, $\operatorname{co-contra-Fr}(\rho)$ назовем классами автоматов, представляемых, соответственно, фрагментами, кофрагментами, контрфрагментами, коконтрфрагментами.

\section{Пример и результат}
Обозначим через $ \mathbb{N}$, $ \mathbb{Z}$, $\mathbb{P}$ множества натуральных, целых и простых чисел, $ \mathbb{Z}^+=\{0,1,2,\ldots\}$.

\begin{figure}[ht]
	\centering
  \includegraphics[width=0.5\textwidth]{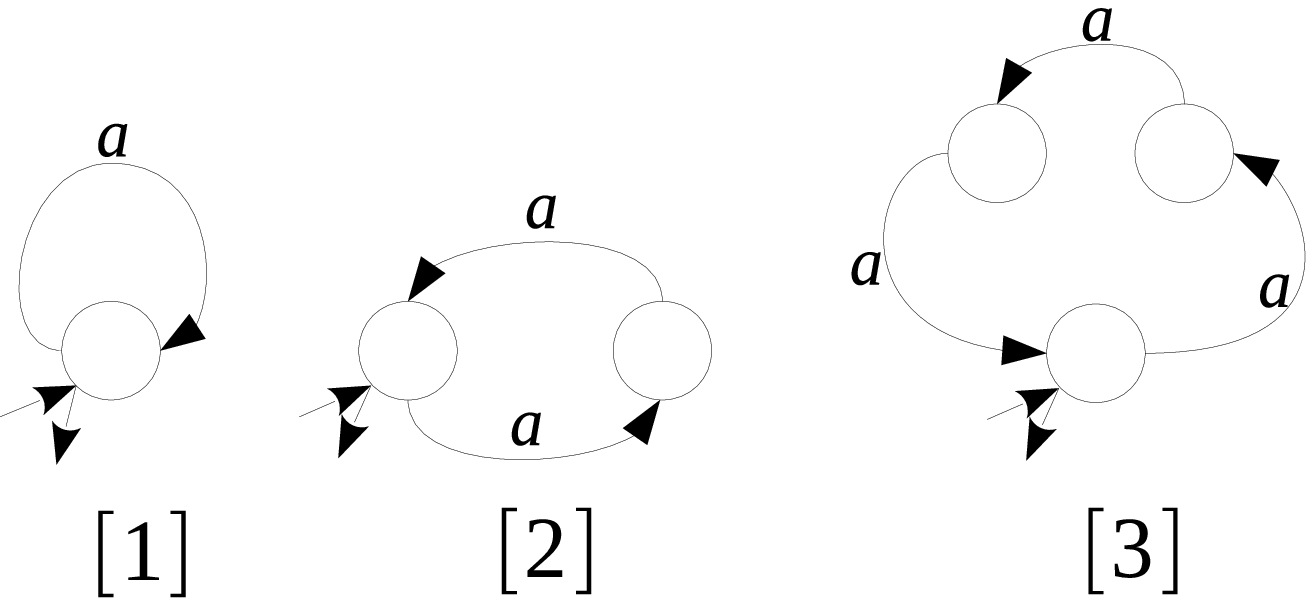}
	\caption{Автоматы $[1]$, $[2]$, $[3]$.}
	\label{fig:123}
\end{figure}

Пусть автомат $[n]=(\{0,1,...,n-1\},\{a\},\delta,0,0)$ состоит из множества состояний $\{0,1,...,n-1\}$,  входного алфавита $\{a\}$, начального состояния $0$, заключительного состояния $0$ и функции переходов $\delta$ такой, что $\delta(i,a)=i+1(\mod n)$ и $\delta(i,a^m)=i+m(\mod n)$. Из определения следует, что автомат $[n]$ допускает язык $\{a^{kn}|k\in\mathbb{Z}^+\}$. 
Через $[0]=\{\{0,1,2,\ldots\},\{a\},\delta,0,0\}$ обозначим автомат, в котором $\delta(n,a)=n+1$. Язык, допускаемый автоматом $[0]$, состоит из пустого слова.
Обозначим класс автоматов $\{[0],[1],[2],\ldots\}$ через $\mathbb{Z}^+$, который далее будем называть \textit{допустимым} классом. Примеры автоматов $[1]$, $[2]$ и $[3]$ показаны на рис.~\ref{fig:123}.

\begin{figure}[ht]
	\centering
  \includegraphics[width=0.5\textwidth]{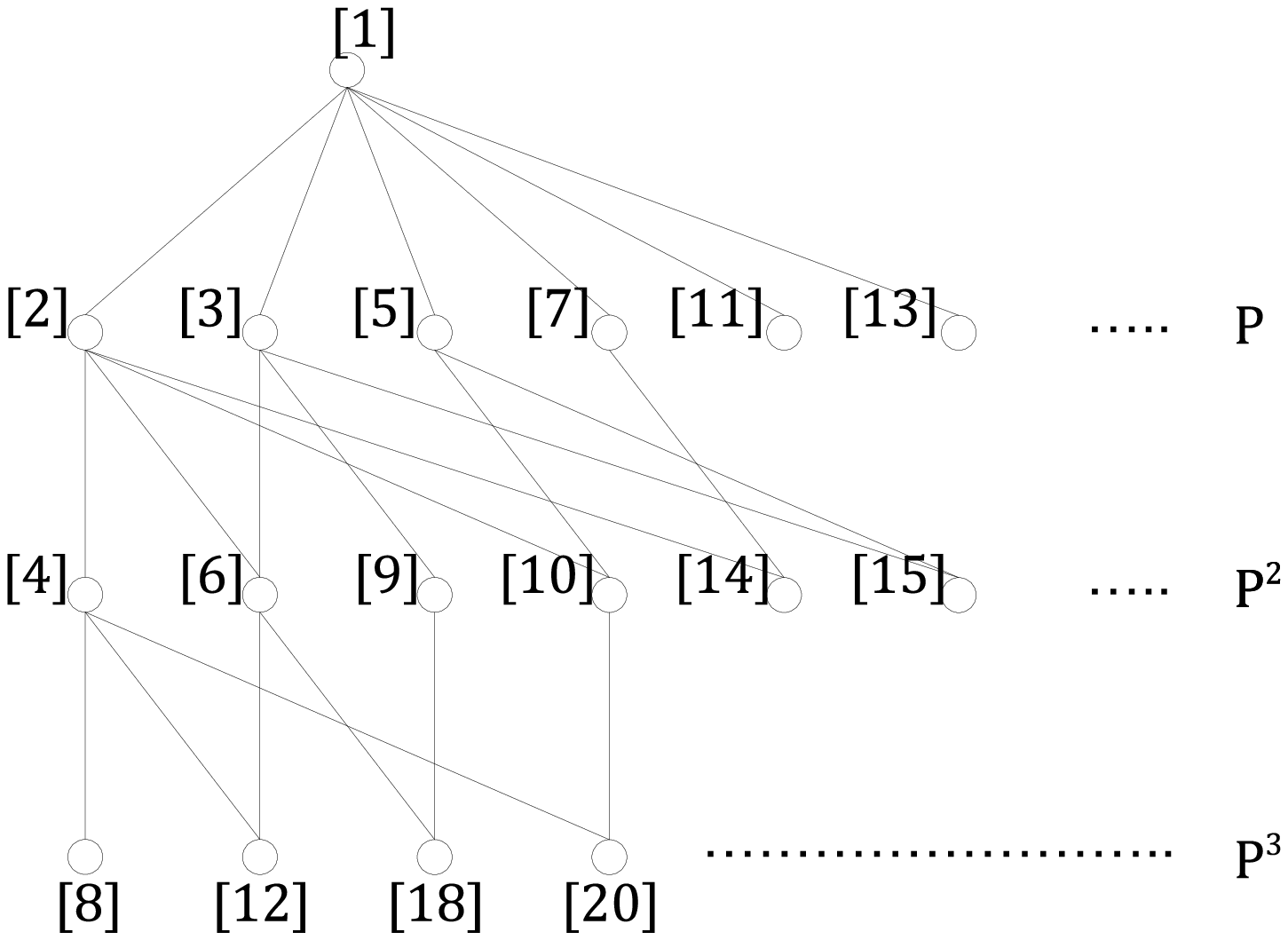}
	\caption{Фрагмент решётки автоматов класса $\Zed^+$.}
	\label{fig:reshetkan}
\end{figure}

Гомоморфизм из автомата $[n]$ в автомат $[m]$ существует тогда и только тогда, когда число $m$ делит число $n$. 
Введем операции $\lor$ и $\land$ на автоматах класса $\mathbb{N}$: $[n]\lor [m]=\gcd(n,m)$ ($\gcd$ -- наибольший общий делитель) наибольший по числу состояний автомат, в который есть гомоморфизм как из $[n]$, так и из $[m]$; $[n]\land [m]=\operatorname{lcm}(n,m)$ ($\operatorname{lcm}$ -- наименьшее общее кратное) наименьший по числу состояний автомат, из которого есть гомоморфизм как в $[n]$, так и в $[m]$.  Например, $[n]\lor [1]=[1]$ и $[n]\land [1]=[n]$ для любого $n\in\mathbb{N}$. 
Считаем по определению, что $[n]\lor [0]=[n]$,  $[n]\land [0]=[0]$.  Класс автоматов $\mathbb{Z}^+$ является полной решёткой без дополнений, фрагмент которой проиллюстрирован на рис.~\ref{fig:reshetkan}.

Слово $a^n$ длины $n$ в алфавите $\{a\}$ далее обозначаем числом $n$ и будем говорить, что автомат допускает не множество слов, а множество чисел. Для примера, автомат $[n]$ допускает  множество чисел $\{0,n,2n,3n,\ldots\}=\{nk|k\in\mathbb{Z}^+\}$.

Для класса автоматов $\mathbb{Z}^+$, когда $\rho\subseteq \mathbb{Z}^+\times\mathbb{Z}^+$, правила конгруэнтного замыкания выглядят так (мы пишем $n=m$ вместо $(n,m)\in\rho$):
\[
\frac{ 
\begin{array}{cc}
n=m, & m=l \end{array}
}{n=l}, 
\frac{n=m}{m=n}, \frac{}{n=n},\frac{n=m}{n+l=m+l}.\] 

Поскольку автоматы класса $\mathbb{Z}^+$ имеют циклическую структуру и начальное состояние совпадает с заключительным, к указанным правилам замыкания $[\cdot]$ добавим правило:
\[
\frac{n=m}{|n-m|=0},
\] 
где $|n-m|$ -- абсолютное значение числа $n-m$.  Эти пять правил образуют правила $\mathbb{Z}^+$-допустимого замыкания. Результатом замыкания произвольного отношения $\rho\subseteq\mathbb{Z}^+\times\mathbb{Z}^+$ всегда будет автомат класса $\mathbb{Z}^+$. Очевидно, что равенство $n=0$ является определяющим соотношением автомата $[n]$.

Примеры представлений для автоматов из класса допустимых автоматов $\mathbb{Z}^+$:
$$
\operatorname{co-Fr}(1=0)\cap \operatorname{contra-Fr}(1=0) = \mathbb{P} = \{[2],[3],[5],[7],\ldots\}
$$
$$
\operatorname{co-Fr}(2=0)\cap \operatorname{contra-Fr}(1=0) = \{[2]\}
$$
$$
\operatorname{co-Fr}(1=0)\cap\operatorname{contra-Fr}(1=0)\cap
\operatorname{Fr}(6=0)\cap\operatorname{co-contra-Fr}(6=0)=\{[2],[3]\}
$$

Очевидно, $\operatorname{Fr}([0])=\operatorname{co-Fr}([1])=\mathbb{Z}^+$, то есть  фрагмент $0=0$ и кофрагмент $1=0$ не накладывают ограничений на представление автоматов. Рассмотрим подкласс $\operatorname{contra-Fr}([1])$ класса $\mathbb{Z}^+$. Очевидно, что $\operatorname{contra-Fr}([1])=\mathbb{Z}^+-\{[1]\}=\{[2],[3],[4],...,\}\cup\{[0]\}$. Этот класс не является решеткой.  
Таким образом, верна теорема:

\begin{theorem}
Подкласс класса $\mathbb{Z}^+$, задаваемый фрагментом и/или кофрагментом,  является полной решёткой. Подкласс, задаваемый контрфрагментом или коконтрафрагментом, является полурешёткой. Подкласс задаваемый, контрфрагментом и коконтрфрагментом, не является в общем случае ни решёткой, ни полурешёткой.
\end{theorem}

\end{document}